\documentclass[aps,prc,twocolumn,showpacs,floatfix,nofootinbib,superscriptaddress,amsmath,amssymb,longbibliography]{revtex4-1}

\usepackage{graphicx}
\usepackage{epsfig}
\usepackage{bm}
\usepackage{color}
\usepackage{float}
\usepackage{dcolumn}
\usepackage{multirow} 
\usepackage{hyperref}

\usepackage{amsmath}

\graphicspath{{.}{../}}

\newcommand{\beq}{\begin{equation}}
\newcommand{\eeq}{\end{equation}}
\newcommand{\beqa}{\begin{eqnarray}}
\newcommand{\eeqa}{\end{eqnarray}}

\begin{document}
\title{Optical potential from first principles}
\thanks{This manuscript has been authored by UT-Battelle, LLC under
  Contract No. DE-AC05-00OR22725 with the U.S. Department of
  Energy. The United States Government retains and the publisher, by
  accepting the article for publication, acknowledges that the United
  States Government retains a non-exclusive, paid-up, irrevocable,
  world-wide license to publish or reproduce the published form of
  this manuscript, or allow others to do so, for United States
  Government purposes. The Department of Energy will provide public
  access to these results of federally sponsored research in
  accordance with the DOE Public Access
  Plan. (http://energy.gov/downloads/doe-public-access-plan).}

\author{J.~Rotureau}

\affiliation{NSCL/FRIB Laboratory, Michigan State University, East
  Lansing, Michigan 48824, USA}
\affiliation{JINPA, Oak Ridge National Laboratory, Oak Ridge, TN 37831, USA}

\author{P.~Danielewicz}

\affiliation{NSCL/FRIB Laboratory, Michigan State University, East
  Lansing, Michigan 48824, USA}

\affiliation{Department of Physics and Astronomy, Michigan State
  University, East Lansing, MI 48824-1321}

\author{G.~Hagen}
\affiliation{Physics Division, Oak Ridge National Laboratory,
Oak Ridge, TN 37831, USA} 
\affiliation{Department of Physics and Astronomy, University of Tennessee,
Knoxville, TN 37996, USA}

\author{F. M.~Nunes}
\affiliation{NSCL/FRIB Laboratory, Michigan State University, East
  Lansing, Michigan 48824, USA}
\affiliation{Department of Physics and Astronomy, Michigan State
  University, East Lansing, MI 48824-1321}

\author{T.~Papenbrock}
\affiliation{Physics Division, Oak Ridge National Laboratory, Oak
  Ridge, TN 37831, USA}
\affiliation{Department of Physics and Astronomy, University of
  Tennessee, Knoxville, TN 37996, USA}

\begin{abstract}
  We develop a method to construct a microscopic optical potential
  from chiral interactions for nucleon-nucleus scattering. 
  The optical potential is constructed by combining the
  Green's function approach with the coupled-cluster method. To deal
  with the poles of the Green's function along the real energy axis we employ
  a Berggren basis in the complex energy plane combined with the
  Lanczos method. Using this approach, we perform a proof-of-principle
  calculation of the optical potential for the elastic neutron
  scattering on $^{16}{\rm O}$. For the computation of the
  ground-state of $^{16}{\rm O}$, we use the coupled-cluster method in
  the singles-and-doubles approximation, while for the $A=\pm 1$
  nuclei we use particle-attached/removed equation-of-motion method
  truncated at two-particle-one-hole and one-particle-two-hole
  excitations, respectively. We verify the convergence of the optical
  potential and scattering phase shifts with respect to the
  model-space size and the number of discretized complex continuum
  states. We also investigate the absorptive component of the optical
  potential (which reflects the opening of inelastic channels) by computing its
  imaginary volume integral  and find an
  almost negligible absorptive component at low-energies. To shed
  light on this result, we computed excited states of $^{16}{\rm O}$
  using equation-of-motion coupled-cluster method with singles-and-doubles
  excitations and we found no low-lying excited states below
  10~MeV. Furthermore, most excited states have a dominant
  two-particle-two-hole component, making higher-order particle-hole
  excitations necessary to achieve a precise description of these
  core-excited states. We conclude that the reduced absorption at low-energies can be attributed to the lack of correlations coming from the low-order cluster
  truncation in the employed coupled-cluster method.
\end{abstract}
\maketitle 
\section{Introduction}
Nuclear reactions are the ubiquitous experimental tool to study atomic
nuclei.  While many astrophysically relevant reactions proceed at
relatively low energies $E<1$~MeV~\cite{ReactionsBook},
in the laboratory, these reactions are often studied indirectly with beams at higher energy ($\geq$5 MeV/u).
One of the most important open questions currently being explored today in our field concerns the astrophysical site for the r-process, the process that gave rise to about half of the heavy elements in our planet. In order to perform simulations of neutron star mergers or supernovae explosions (the two possible sites under consideration), neutron capture rates are needed on rare isotopes of nuclei as heavy as Uranium \cite{ReactionsBook}. Despite all the effort  with {\it ab initio} approaches to nuclear reactions, which include the study of elastic scattering~\cite{nollett2007,quaglioni2008,hagen2012c,hupin2013,elhatisari2015,mazur2015},
transfer~\cite{quaglioni2012}, photo
reactions~\cite{gazit2006,efros2007,bacca2014b}, and capture
reactions~\cite{girlanda2010,marcucci2013}, only selected nuclei and specific reaction channels can be addressed with the various ab-initio methods in the market  (Refs.~\cite{bacca2014,navratil2016} for recent
reviews). 

A more general approach to  reactions involving heavier nuclei is based on a reduction of the many-body picture to a few-body one, where only the most relevant degrees of freedom are retained \cite{ReactionsBook}. In such approaches one introduces effective interactions (the so-called optical potentials) between the clusters considered. Traditionally these interactions have been constrained by data, particularly using data on $\beta$-stable isotopes~\cite{KD,pheno}. Clearly, the application of these global parameterizations to exotic regions of the nuclear chart is unreliable and has uncontrolled uncertainties. It is critical for progress in the field of reactions that these effective interactions be connected to the underlying microscopic theory so that extrapolations to exotic regions can be better understood.

In most cases,  phenomenological optical potentials are made local for simplicity. 
We know based on the Feshbach projection formalism that, in its most general form, the
microscopic optical potential should be complex, non-local and energy
dependent~\cite{Feshbach1,Feshbach2}. Recently, a series of studies has shown that nonlocality can affect transfer reaction observables
(e.g. \cite{Titus_prc2014,Ross_prc2015,Titus_prc2016}) and it is expected that it can equally affect other reaction channels. So far we have not been able to identify an experimental method to constrain nonlocality. It is essential that  microscopic theories provide guidance on this aspect of the optical potential.

The goal of this work is to provide a proof-of-principle for a new method to compute nuclear optical potentials from ab-initio many-body coupled cluster calculations. It is the first of a series of studies that aims at constructing an optical potential rooted in the underlying microscopic formulation of the problem, potentials which can then be incorporated, consistently with other ingredients, into the general few-body formalism. In an approach based on Feshbach projection operators, the optical 
potential is the self-energy term in the Dyson equation~\cite{capuzzi}.  
Semiphenomenological optical potential have been obtained using approximation of the self-energy at the  Brueckner-Hartree-Fock  level~\cite{jeuk}. For the scattering
of nucleons at high energy ($\geq$ 100 MeV) optical potential can be derived with the multiple scattering formalism \cite{kerman}.
More recently, the solution of the Dyson equation by
self-consistent Green's function methods has been used to compute optical potentials~\cite{dickhoff2004,barbieri2005,DOM}.  In this
paper, we compute the Green's function directly following the coupled cluster method~\cite{kuemmel1978,hagen2014}, thus circumventing the usual self-consistency approach. 
The self-energy can then be determined by inverting the Dyson equation. The key elements in our
approach to compute the Green's function are: i) an analytical
continuation in the complex energy plane based on a Berggren basis
consisting of bound, resonant, and non-resonant scattering
states~\cite{berggren1968,michel2002,idbetan2002,hagen2004,hagen2006},
and ii) a generalized non-symmetric Lanczos method \cite{cullum98} that
allows us to write the Green's function as a continued fraction
\cite{dagotto1994,hallberg1995,efros2007,bacca2014b}. The first of these two elements is essential because it allow us
to properly deal with the poles of the Green's function along the real
energy axis, and obtain numerically stable Green's functions and
optical potentials. The second element is essential to make the problem computationally feasible.
In this work we demonstrate that optical potentials, converged with respect to the models space, can indeed be determined from the Green's functions generated from coupled cluster many-body calculations. We note that the computation of
Green's functions with the coupled-cluster method is well established
in quantum chemistry~\cite{gf_cc1,gf_cc2,gf_cc3}, and that very
recently this approach has also been used to extract the optical
potential~\cite{bhaskaran2016}. Our approach is similar to that effort, but applied to nuclear many-body problem. 

This paper is organized as follows. In Sec.~(\ref{sec_formalism}) we
introduce the formalism of the Green's function and the
coupled-cluster method along with the Berggren basis and discuss the
application of the Lanczos method for the numerical calculations of
the Green's function.  In Sec.~\ref{secres}, we show an application
for the elastic scattering on ${\rm ^{16}O }$ and discuss the
results. Finally, we will conclude and discuss future possible
applications in Sec.~\ref{conclusion}.
\section{Formalism} \label{sec_formalism}
\subsection{The single-particle Green's function} \label{spgf}
The single-particle Green's function of an $A$-nucleon system
has matrix elements
\begin{eqnarray}
G(\alpha,\beta,E)&=&\langle \Psi_{0}|a_{\alpha}\frac{1}{E-(H-E^{A}_{gs})+i \eta} a^{\dagger}_{\beta}|\Psi_0\rangle \nonumber \\
&+& \langle \Psi_{0}|a^{\dagger}_{\beta}\frac{1}{E-(E^{A}_{gs}-H)-i \eta} a_{\alpha}|\Psi_0\rangle .
\label{gf1}
\end{eqnarray}
Here, $\alpha$ and $\beta$ denote single-particle states, and
$|\Psi_0\rangle$ is the ground state of the $A$-body system with
energy $E^{A}_{gs}$. As usual, the parameter $\eta\ge 0$ is such that
$\eta \rightarrow 0$ at the end of the calculation.  The operators
$a^{\dagger}_\alpha$ and $a_\beta$ create and annihilate a fermion in
the single-particle state $\alpha$ and $\beta$, respectively, and are
shorthands for the quantum numbers $\alpha=(n,l,j,j_z,\tau_z)$.  Here, $n,l,j,j_z,\tau_z$ label the radial quantum number, the orbital
angular momentum, the total orbital momentum, its projection on the
$z-$ axis, and the isospin projection, respectively.  The intrinsic
Hamiltonian $H$ is
\begin{eqnarray}
H=\sum^{A}_{i=1}\frac{\vec{p_i}^2}{2m}
-\frac{\vec{P}^2}{2mA} +\sum_{i<j} V_{ij} .
\label {hami}
\end{eqnarray}
Here, $\vec{p}_i$ is the momentum of the nucleon $i$ of mass $m$ and
$\vec{P}=\sum_{i=1}^A\vec{p}_i$ is the momentum associated with the center of mass motion.  We limit ourselves to a two-body
interactions $V_{ij}$ and neglect contributions from three-nucleon
forces. It is useful to rewrite the Hamiltonian as
\begin{eqnarray}
H=\sum^{A}_{i=1}\frac{\vec{p}_i^2}{2m} \left (1-\frac{1}{A}\right)   +\sum_{i<j} \left ( V_{ij} -\frac{\vec{p}_i\vec {p}_j}{mA} \right ) , \label {hami2}
\end{eqnarray}
separating one-body and two-body contributions. In what follows, we
take the single-particle states from the Hartree-Fock (HF) basis. We
recall that the HF basis is an excellent starting point for
coupled-cluster calculations and that the HF Green's function
\begin{eqnarray}
  \label{HFGF}
G^{(0)}(\alpha,\beta,E)&=&\langle \Phi_{0}|a_{\alpha}\frac{1}{E-(H_0-E^{A}_{gs,0})+i \eta} a^{\dagger}_{\beta}|\Phi_0\rangle \nonumber \\
&+& \langle \Phi_{0}|a^{\dagger}_{\beta}\frac{1}{E-(E^{A}_{gs,0}-H_0)-i \eta} a_{\alpha}|\Phi_0\rangle
\end{eqnarray}
is a first order approximation to the Green's function (\ref{gf1}). 
 In Eq.~(\ref{HFGF}) $H_0$ is the HF potential,
$|\Phi_{0}\rangle$ the HF reference state of
the $A$-nucleon system and $E^{A}_{gs,0}$ the corresponding energy. As
the single-particle states $\alpha, \beta$ are given by the HF basis,
Eq.~(\ref{HFGF}) can be written as
\begin{eqnarray}
G^{(0)}(\alpha,\beta,E)=\delta_{\alpha,\beta} \left [ \frac{\Theta(\alpha-F)}{E-\varepsilon_{\alpha}+i\eta}
+\frac{\Theta(F-\alpha)}{E-\varepsilon_{\alpha}-i\eta}\right ].
\end{eqnarray}
Here, $\varepsilon_{\alpha}$ is the single-particle energy associated with
$|\alpha\rangle$ and $\Theta$ the unit step function. For a single-particle state
$\alpha$ above the occupied shells in the HF approximation,
$\Theta(\alpha-F)=1$, whereas $\Theta(\alpha'-F)=0$ for $\alpha'$ below
the Fermi level.

The Green's function fulfills the Dyson equation 
\begin{eqnarray}
G(\alpha,\beta,E)&=&G^{(0)}(\alpha,\beta,E) \nonumber \\
&+&\sum_{\gamma,\delta}G^{(0)}(\alpha,\gamma,E)\Sigma^{*}(\gamma,\delta,E)G(\delta,\beta,E) .
\label{dys}
\end{eqnarray}
Here, $\Sigma^{*}(\gamma,\delta,E)$ is the self energy, which can be obtained 
from the inversion of  Eq.~(\ref{dys}):
\begin{eqnarray}
\Sigma^{*}(E)=[G^{(0)}(E)]^{-1}-G^{-1}(E) .
\end{eqnarray}
To obtain the optical potential we introduce the quantity
\begin{eqnarray}
\Sigma'\equiv \Sigma^*+U, \label{optdef}
\end{eqnarray}
where $U$  is the HF potential. For $E\geq E^{A}_{gs}$, $\Sigma'$ in
Eq.~(\ref{optdef}) corresponds to the optical potential for the
elastic scattering from the $A$-nucleon ground state~\cite{capuzzi,dickhoff}.
We are interested in the scattering amplitude
\begin{eqnarray}
\xi_{E^+}({\bf r})=\langle \Psi_{0}|a_{{\bf r}}|\Psi_{E+}\rangle 
\end{eqnarray}
where $|\Psi_{E^+}\rangle$ is the elastic scattering state of a nucleon
on the target with the energy $E^+=E-E^{A}_{gs}$ and $a_{\bf{r}}$ the annihilation operator of a particle at the position ${\bf{r}}$.  The scattering
amplitude $\xi^c_{E^+}(r)$ is the solution of the Schr\"odinger
equation containing the optical potential 
\begin{eqnarray}
-\frac{\hbar^2}{2\mu}\nabla^2\xi({\bf r})+\int d{ \bf r'} 
\Sigma'( {\bf r},{\bf r'},E^+) \xi({\bf r'})=E^+ \xi({\bf r}) .
\label{schro}
\end{eqnarray}
where $\mu$ is the reduced mass of the nucleus-nucleon system.
For simplicity, we
suppressed any spin and isospin labels.  The optical potential is
non-local, energy-dependent and complex \cite{dickhoff} and, for
$E^+\geq 0$, its imaginary component describes the loss of flux
due to absorption. Similarly, the overlap $\xi_{n}({\bf r})=\langle
\Psi_{0}|a_{\bf{r}}|\Psi^{A+1}_{n}\rangle$ for a bound state
$|\Psi^{A+1}_{n}\rangle$ of energy $E_n^{A+1}$ in the $A+1$ system,
fulfills the Schr\"odinger equation with the optical potential at the
discrete energy $E_n=E_n^{A+1}-E^{A}_{gs}$.

In this paper, we construct the optical potential by an inversion of
the Dyson equation (\ref{dys}) after a direct computation of the Green's
function (\ref{gf1}) following the coupled-cluster method \cite{hagen2014}.
In the following section, we present the main steps involved in the
computation of the Green's function in our approach.

\subsection{Green's function from coupled-cluster method}
The HF reference state for the nucleus consisting of $A$ nucleons is
\begin{eqnarray}
|\Phi_{0}\rangle=\Pi_{i=1}^{A}a^{\dagger}_i|0\rangle .
\end{eqnarray}
In coupled-cluster theory, see Refs.~\cite{bartlett2007,hagen2014} for
details, the ground state is represented as
\begin{eqnarray}
|\Psi_{0}\rangle=e^T|\Phi_{0}\rangle \label{cc1} , 
\end{eqnarray}
and $T$ denotes the cluster operator
\begin{eqnarray}
T &=& T_1+T_2+\dots \nonumber \\
&=& \sum_{i,a}t_i^a a^{\dagger}_a a_i+\frac{1}{4}\sum_{ijab}t_{ij}^{ab}t_{ijab}a^{\dagger}_aa^{\dagger}_ba_ja_i +\ldots .
\label{t_cluster}
\end{eqnarray}
We note that $T_1$ and $T_2$ induce $1p$-$1h$ and $2p$-$2h$
excitations of the HF reference, respectively. Here and in what
follows, the single-particle states $i,j,...$ refer to hole states
occupied in the reference state $|\Phi_{0}\rangle$ while $a,b,...$
denote valence states above the reference state.  In practice, the
expansion (\ref{t_cluster}) is truncated. In the coupled cluster with
singles and doubles (CCSD) all operators $T_i$ with $i>2$ are
neglected. In that case, the ground-state energy and the amplitudes
$t_i^a, t_{ij}^{ab}$ are obtained by projecting the state (\ref{cc1}) on the
reference state and on all $1p$-$1h$ and $2p$-$2h$ configurations for which
\begin{eqnarray}
  \label{ccsd}
  \nonumber
  \langle \Phi_0|\overline{H}|\Phi_0\rangle&=&E ,  \\
  \nonumber
  \langle \Phi_i^a|\overline{H}|\Phi_0\rangle&=&0 ,  \\
 \langle \Phi_{ij}^{ab}|\overline{H}|\Phi_0\rangle&=&0 .
\end{eqnarray}
Here,
\begin{eqnarray}
  \overline{H}&\equiv& e^{-T}He^T \nonumber\\
  &=& H + \left[H,T\right] +{1\over 2!}\left[\left[H,T\right],T\right] + \ldots
\end{eqnarray}
denotes the similarity transformed Hamiltonian and it can be computed
systematically  via the Baker-Campbell-Hausdorff expansion. For two-body
forces and in the CCSD approximation, this expansion actually terminates at
fourfold nested commutators.

The CCSD equations~(\ref{ccsd}) show that the CCSD ground state
is an eigenstate of the similarity-transformed Hamiltonian
in the space of $0p$-$0h$, $1p$-$1h$, $2p$-$2h$
configurations. The transformed Hamiltonian is not Hermitian because
the operator $e^{T}$ is not unitary. As a consequence, $\overline{H}$
has left- and right-eigenvectors which constitute a bi-orthogonal
basis with the corresponding completeness relation
\begin{eqnarray}
\sum_{i}|\Phi_{i,R}\rangle \langle \Phi_{i,L}|=\hat{{1}} .
\end{eqnarray}
The right ground state is the reference state $|\Phi_0\rangle$, while the
left ground-state is given by $\langle \Phi_{0,L} | =
\langle \Phi_0 |(1 + \Lambda)$ where $\Lambda$ is a linear combination
of particle-hole de-excitation operators.

Using the ground state of the similarity-transformed Hamiltonian, we
now can write the coupled cluster Green's function $G^{CC}$ as
\begin{eqnarray}
\lefteqn{G^{CC}(\alpha,\beta,E) \equiv }\nonumber\\
&& \langle \Phi_{0,L}|\overline{a_{\alpha}}\frac{1}{E-(\overline{H}-E^{A}_{gs})+i\eta}\overline{a^{\dagger}_{\beta}}|\Phi_{0}\rangle \nonumber \\
&+&\langle \Phi_{0,L}|\overline{a^{\dagger}_{\beta}}\frac{1}{E-(E^{A}_{gs}-\overline{H})-i\eta}\overline{a_{\alpha}}|\Phi_{0}\rangle .
\label{gfcc}
\end{eqnarray}
Here, $\overline{a_{\alpha}}=e^{-T}a_{\alpha}e^T$ and
$\overline{a^{\dagger}_{\beta}}=e^{-T}a^{\dagger}_{\beta}e^T$ are the
similarity-transformed annihilation and creation operators,
respectively, and the Baker-Campbell-Hausdorff expansion yields the relations
\begin{eqnarray}
\overline{a_{\alpha}}&=&a_{\alpha}+[a_{\alpha},T] , \\
\overline{a^{\dagger}_{\beta}}&=&a^{\dagger}_{\beta}+[a^{\dagger}_{\beta},T] .
\end{eqnarray}
We note that the truncation of the cluster operator $T$ is reflected
in the expression of the coupled-cluster Green's function
(\ref{gfcc}), and if all excitations up to $Ap$-$Ah$ were taken into
account in the expansion (\ref{t_cluster}), the Green's function
(\ref{gfcc}) would be identical to (\ref{gf1}).

One might be tempted to use the completeness relations for the $A \pm 1$ 
systems to obtain the Lehmann representation of the Green's function
\begin{eqnarray}
\lefteqn{G^{CC}(\alpha,\beta,E)=} \nonumber \\
&& \sum_{i} \frac{\langle \Phi_{0,L}|\overline{a_{\alpha}}|\Phi^{A+1}_{i}\rangle \langle \Phi^{A+1}_{i}|\overline{a^{\dagger}_{\beta}}|\Phi_{0}\rangle}{E-(E^{A+1}_{i}-E^{A}_{gs})+i\eta} \nonumber \\
&+&\sum_j\frac{\langle \Phi_{0,L}|\overline{a^{\dagger}_{\beta}}|\Phi^{A-1}_{j}\rangle \langle \Phi^{A-1}_j|\overline{a_{\alpha}}|\Phi_{0}\rangle}{E-(E^{A}_{gs}-E^{A-1}_j)-i\eta} .
\label{gfcc2}
\end{eqnarray}
Here, $|\Phi^{A+1}_i\rangle$ ($|\Phi^{A-1}_j\rangle$) is an eigenstate
of $\overline{H}$ for the $A+1$ ($A-1$) system with energy $E^{A+1}_i$
($E^{A-1}_j$).  To simplify the notation, the completeness relations
are written in (\ref{gfcc2}) as discrete summations over the states in
the $A\pm 1$ systems. In principle, the Green's function (\ref{gfcc2})
could be obtained by calculating the spectrum of the $A\pm 1$ systems
using the particle-attached equation-of-motion (PA-EOM) and
particle-removed equation-of-motion (PR-EOM) coupled-cluster
methods~\cite{gour2006}. However in practice, this approach is
difficult to pursue as the sum over all states also involves eigenstates in the
continuum.  To avoid this problem, we return to the expression
in Eq.~(\ref{gfcc}) and use the Lanczos technique \cite{efros2007,bacca2014b}
for its computation.

\subsection{Lanczos method}
\label{seclanc}
In this section, we describe the calculation of the coupled-cluster Green's function
(\ref{gfcc}) using the Lanczos method
\cite{dagotto1994,hallberg1995,efros2007,bacca2014b}. To simplify the
notation we introduce the following shorthands
\begin{eqnarray}
|v^{A+1}_{\beta}\rangle&\equiv&\overline{a^{\dagger}_{\beta}}|\Phi_{0}\rangle , \\
\langle u^{A+1}_{\alpha}|&\equiv&\langle \Phi_{0,L}|\overline{a_{\alpha}} , \\
|v^{A-1}_{\alpha}\rangle&\equiv&\overline{a_{\alpha}}|\Phi_{0}\rangle , \\
\langle u^{A-1}_{\beta}|&\equiv&\langle \Phi_{0,L}|\overline{a^{\dagger}_{\beta}} ,
\end{eqnarray}
and write the Green's function as 
\begin{eqnarray}
\lefteqn{ G^{CC}(\alpha,\beta,E)=}\nonumber \\
&& \langle u_{\alpha}^{A+1} |\frac{1}{E-(\overline{H}-E^{A}_{gs})+i\eta}| v^{A+1}_{\beta} \rangle \nonumber  \\
&+&\langle u^{A-1}_{\beta}|  \frac{1}{E-(E^{A}_{gs}-\overline{H})-i\eta}|v^{A-1}_{\alpha}\rangle .
\label{gfcc3}
\end{eqnarray}
For a truncation of $T$ at the $np$-$nh$ level, the states
$|v^{A+1}_{\beta}\rangle$ and $\langle u^{A+1}_{\alpha}|$ belong to
the vector space ${\cal V}^{A+1}$ spanned by the states built from
$1p$-$0h$,...,$np$-$(n-1)h$ excitations of the reference state
$|\Phi_0\rangle$. Similarly, the states $|v^{A-1}{\alpha}\rangle$ and
$\langle u^{A-1}_{\beta}|$ belong to 
the vector space ${\cal V}^{A-1}$
spanned by $0p$-$1h$,..,$(n-1)p$-$nh$ excitations of the reference state.
Introducing $|X_{\beta}\rangle$ and $|Y_{\alpha}\rangle$ defined as
\begin{eqnarray}
\left [z_+-\overline{H}\right]|X_{\beta}\rangle&\equiv&|v^{A+1}_{\beta}\rangle ,  \label{eqx} \\
\left [z_-+\overline{H}\right]|Y_{\alpha}\rangle  &\equiv&|v^{A-1}_{\alpha}\rangle  , \label{eqy}
\end{eqnarray}
with $z_+\equiv E+E^{A}_{gs}+i\eta$ and $z_-\equiv
E-E^{A}_{gs}-i\eta$, we can write:
\begin{eqnarray}
  G^{CC}(\alpha,\beta,E)=
  \langle u_{\alpha}^{A+1}|X_{\beta}\rangle+\langle u^{A-1}_{\beta}|Y_{\alpha}\rangle .
  \label {gfcc4}
\end{eqnarray}
This matrix element of the Green's function is calculated by solving
the systems of linear equations (\ref{eqx}) and (\ref{eqy}) in the
Lanczos basis.  The advantage of working in the Lanczos basis is
twofold. First, the actual dimensions of the linear systems (defined
by the number of Lanczos vectors $N_{\rm lanc}$) needed to reach
convergence, are much smaller than the dimension of the full space
${\cal V}^{A+1}$ and ${\cal V}^{A-1}$. Second, the resolution has to
be done only once for all energies $E$.

Let us now focus on the first term on the right hand side of (\ref{gfcc4}), i.e.  the term
associated with the particle part of the Green's function.  Starting
with the normalized states $\frac{|v^{A+1}_{\beta}\rangle}{N_0}$ and
$\frac{\langle u_{\alpha}^{A+1}|}{N_0}$ (where the norm is
$N_0=\sqrt{\langle u_{\alpha}^{A+1}|v^{A+1}_{\beta}\rangle }$) as
right and left Lanczos pivots, we construct iteratively a set of
$N_{\rm lanc}$ pairs of Lanczos vectors. By construction, $\overline{H}$
is conveniently represented in the Lanczos basis as a tridiagonal
matrix:
\[\begin{bmatrix}
a_0 & b_0 & 0 & 0 & \ldots \\
b_0  & a_1 &b_1   &0 & \ldots  \\
0  &  b_1 &a_2  &b_2 & \ldots \\
\vdots & \vdots & \vdots & \vdots & \ddots
\end{bmatrix}\]
Using the Cramer's rule for the resolution of linear systems, one can
then show that $\langle u_{\alpha}^{A+1}|X_{\beta}\rangle$ is given by
the continued fraction
\begin{eqnarray}
\lefteqn{\langle u^{A+1}|X_{\beta}\rangle=} \nonumber \\
&& \cfrac{N_0}{(z_+-a_0) 
          - \cfrac{b_{0}^2}{(z_+-a_1) 
          - \cfrac{b^2_1}{(z_+-a_2) -\ldots } } } .
\end{eqnarray}
As it is clear from the expression above, one just needs to solve the
linear system~(\ref{eqx}) only once in order to calculate $\langle
u^{A+1}|X_{\beta}\rangle$ for any value of the energy $E$. The
convergence as a function of $N_{\rm lanc}$ is quickly reached as we will
show in Sec.~(\ref{secres}). The calculation of the second term in
(\ref{gfcc4}), i.e. the hole part of the Green's function, proceeds in
a similar manner.

\subsection{Berggren basis} \label{berg} 
Ultimately we want to compute
the optical potential describing scattering processes at arbitrary
energies. However, as $\eta \rightarrow 0$, the coupled-cluster Greens' function in
Eq.~(\ref{gfcc2}) has poles at energies $E=(E^{A+1}_i-E^{A}_{gs})$ which make the numerical calculation
unstable.
There have been
various proposed solutions to this problem, such as using a complex
scaling technique
\cite{suzuki2005,kruppa2007,carbonell2014,papadimitriou2015}, or carrying calculations at finite values of $\eta$
and extrapolating to $\eta \rightarrow 0$  \cite{braun2014}. 
Another (phenomenological) approach to this
problem is to employ a finite energy dependent width which accounts for damping
and decay processes that are not included 
in the employed theoretical approach~\cite{dickhoff2016}. In this work, we suggest a different
approach based on an analytic continuation of the Green's function in
the complex energy plane using a Berggren basis
\cite{berggren1968,hagen2004}, that includes bound-, resonant, and
discretized non-resonant continuum states. As we will demonstrate
below, by employing the Berggren basis it is possible to obtain stable
numerical results as $\eta \rightarrow 0$.

Thus, the set of HF states includes bound, resonant (when they exist)
and complex-continuum states single-particle states. Accordingly, the
many-body spectrum for the $A+1$ ($A-1$) systems obtained with the
PA-EOM CCSD (PR-EOM CCSD) is composed of bound, resonant and
complex-continuum states. In other words, the poles of the Green's
function [cf Eq.~(\ref{gfcc2})] have either a negative real or complex
energy. As a consequence, as $\eta \rightarrow 0$, the values of the
Green's function matrix elements for $E\geq 0$ smoothly converge to a
finite value. In the case of a real HF basis consisting of bound, and discretized
real energy continuum states, the calculation would become unstable for
small $\eta$ since  the Green's function poles would then be located at real
values of $E$. 

In order to fulfill the Berggren completeness \cite{berggren1968}, the
complex-continuum single-particle states must be located along a
contour $L^+$ in the fourth quadrant of the complex momentum plane
below the resonant single-particle states.  According to the Cauchy
theorem, the precise form of the contour $L^+$ is not important,
provided all resonant states lie between the contour and the real
momentum axis.  The Berggren completeness then reads
\begin{eqnarray}
\sum_{i}|u_i\rangle\langle \tilde{u_i}|+\int_{L^{+}}dk|u(k)\rangle\langle \tilde{u(k)}|= \hat{{1}}, 
\end{eqnarray}
where the discrete states $|u_i\rangle$ correspond to bound and
resonant solutions of the single-particle potential, and $|u(k)\rangle$ are
complex-energy scattering states along the complex-contour $L^+$. In
practise, the integral along the complex continuum is discretized
yielding  a finite discrete basis set. 
\section{Results}\label{secres}
We now present results for the elastic scattering of a neutron on
$^{16}{\rm O}$ . The choice of this problem is motivated by the fact
that ${^{16}\rm O}$ is a doubly magic nucleus and as such can be
computed relatively precisely using the coupled-cluster method.  We will work at
in the CCSD approximation and use the ${\rm {NNLO_{opt}}}$
\cite{ekstrom2013} nucleon-nucleon interaction.  We also want to point
out that we introduced a simplification for the solutions of the
PA-EOM CCSD and the PR-EOM CCSD equations. Instead of solving these
problems with the mass $A+1$ ($A-1$) for the $A+1$ ($A-1$) systems
\cite{hagen2010a}, we have used in all calculations the mass $A=16$. This
introduces a small error (of the order $\sim 1/A$) that is not
relevant in this proof-of-principle calculation.  In principle, the optical potential should be expressed in the neutron-target relative 
coordinates. However, the calculations are performed using the laboratory coordinates
(the Hamiltonian $H$ Eq.~(\ref{hami}) is defined with these coordinates) and we will identify the calculated optical potential
with the optical potential in the relative coordinates. This also introduces a small error of the order $\sim 1/A$.

Table~\ref{tabres} shows the PA-EOM CCSD energies for the low-lying
states in ${\rm ^{17}O}$.  The first two states ($J^{\pi}=5/2^{+},
1/2^{+}$) are bound whereas the second excited state
($J^{\pi}=3/2^{+}$) is resonant. In the computation of these states,
we start the HF calculations in a single-particle basis that employs a
mixed representation of harmonic oscillator states and Berggren states.
We include all harmonic oscillator shells such that $2n+l\leq N_{max}$
and for a given $J^{\pi}$ state in ${\rm ^{17}O}$, we only use Berggren states
for the partial wave $(l,j)$ that couples with the $0^{+}$ gs state in
${\rm ^{16}O }$ to the total angular momentum $J^{\pi}$. For instance,
for the $5/2^+$ ground state in $^{17}$O we use harmonic oscillator
states for all partial waves excepted for the neutron $d_{5/2}$
orbital. We have checked that the results remain unchanged when the Berggren basis is used for multiple orbitals.
 The harmonic oscillator frequency is kept fixed at $\hbar
\omega =20$~MeV.

Energies are practically converged for $N_{\rm max}=14$ at a precision
of few keV for the ground state in $^{17}{\rm O}$ and few tens of keV
for the excited states. We note that due to the non-Hermitian
character of both the CC and the representation of the Hamiltonian in
the Berggren basis, the dependence of the energy with the size of the
model space is not necessarily monotonic. This can be seen, for
instance, in the result for the (complex) energies of the
$J^{\pi}=3/2^{+}$ resonance in $\rm{^{17}O}$.  Table~\ref{tabres} also
shows the CCSD ground-state energy in ${\rm^{16}O}$.

\begin{table}[htb]
	
	\begin{ruledtabular}
		\begin{tabular}{lcccc}
			${N_{\rm max}}$ & ${E(5/2^{+})  }$ & $ { E(1/2^{+}) }$ & ${E(3/2^{+}) }$ & ${E_{gs}(^{16}O)}$ \\[2pt]
				\hline\\[-7pt]
				
                                8  & -4.35 & -2.62 & 2.68-i0.32  & -121.68   \\
                                10 & -4.49 & -2.73 & 2.24-i0.25  & -123.24   \\
                                12 & -4.56 & -2.76 & 2.34-i0.21  & -123.49 \\
                                14 & -4.57 & -2.80 & 2.26-i0.12  & -123.52 \\
                              \end{tabular}
		\end{ruledtabular}

  \caption{PA-EOM CCSD energy of the lowest states in ${\rm ^{17}O}$
    and CCSD ground-state energy in ${\rm ^{16}O}$ with the ${\rm
      NNLO_{opt}}$ \cite{ekstrom2013} interaction.  Results are given
    in MeV. The resonant $J^\pi=3/2^+$ state has a complex energy.}
  \label{tabres}
\end{table}

The calculated ground state of $^{16}$O at the CCSD level is
underbound by about 4~MeV compared to the experimental value at -127.62 MeV, while
CCSD with a perturbative triples correction gives a ground-state
energy of $-130.1$~MeV \cite{ekstrom2013}. The ground-state of
$^{17}{\rm O}$ is found to be overbound by about 0.4~MeV ($E_{\rm
  exp}(5/2^+)=-4.14$~MeV). The first excited state is underbound by
about 0.5~MeV ($E_{\rm exp}(1/2^+)=-3.272$~MeV), and the real part of
the energy of the resonant $J^{\pi}=3/2^{+}$ state is about 1.3 MeV
above the experimental value $E_{\rm exp}(3/2^+)= 0.943
-i0.48$~MeV. One can speculate whether higher order correlations such
as $3p$-$2h$ excitations in the PA-EOM approach, and the neglected
three-nucleon forces will impact these low-lying states in $^{17}$O.
We also remind the reader that we have used $A=16$ in the PA-EOM CCSD
calculations of $^{17}$O which introduce a small error in the
computation of total binding energies. A more significant effect is
seen if one looks at the energies of $^{17}$O with respect to the
ground-state of $^{16}$O, shown in Tab.~\ref{tabres}. In the PA-EOM
CCSD computations of $^{17}$O, the energies are given with respect to
the ground-state of $^{16}$O, i.e. $\omega_\mu^{A+1} = E_\mu^{A+1} -
E_0^*$.  Using $A=17$ for $^{17}$O the ground-state energy $E_0^*$ of
$^{16}$O is computed with the same mass $A = 17$, so in order to get
the correct threshold one needs to add the energy shift $E_0^* - E_0$
where $E_0$ is the ground-state energy of $^{16}$O with $A=16$, this
shift is about $-0.7$~MeV for the states show in in Table~\ref{tabres}
(see e.g. \cite{hagen2010a,hagen2012c} for more details).

\begin{figure}[htb]
\begin{center}
\includegraphics[scale=0.3]{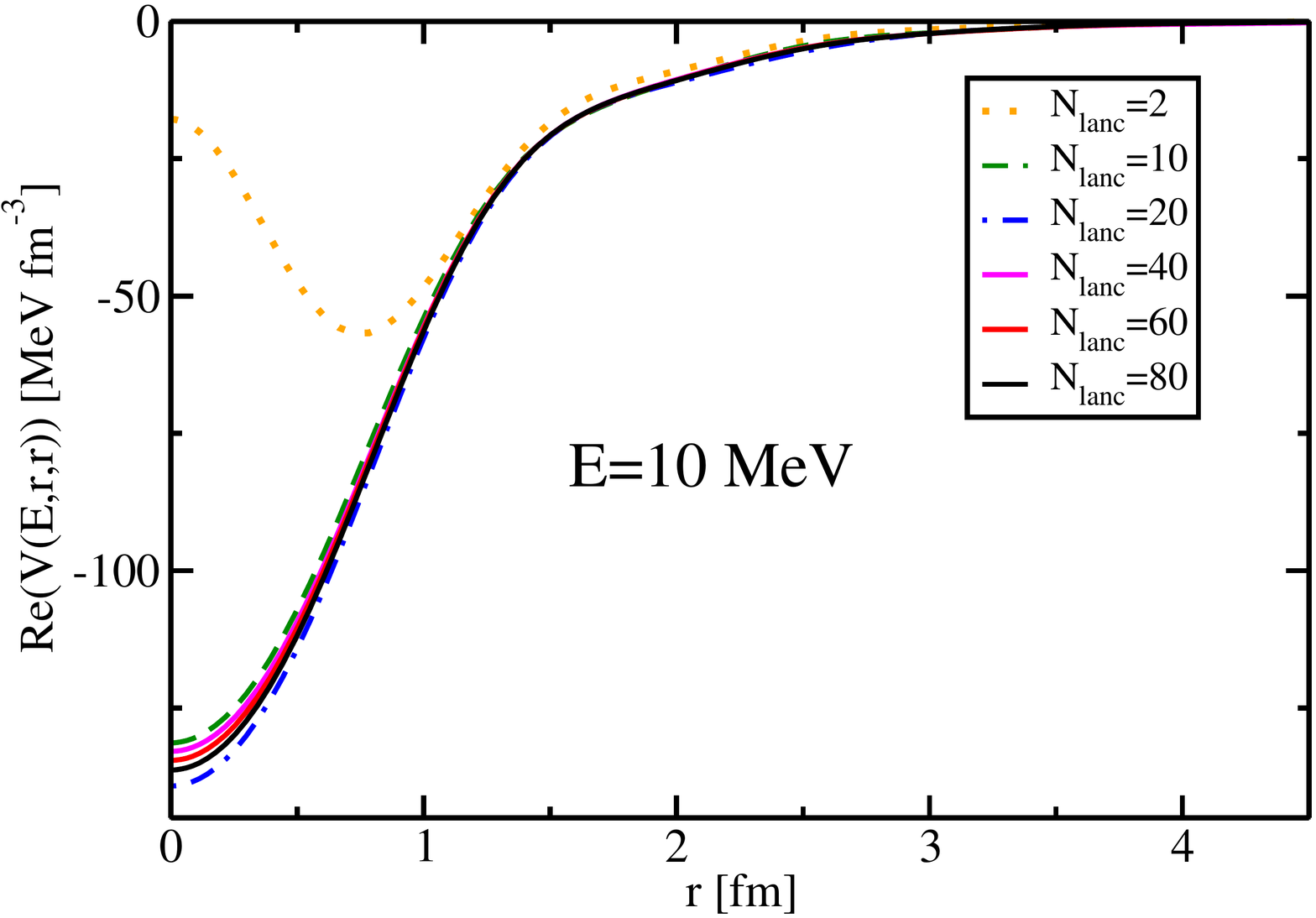}
\caption{Real part of the radial (diagonal) optical potential in the
  neutron $s$-wave at $E=10$~MeV as a function of ${N_{\rm lanc}}$ the
  number of Lanczos iteration. Calculations were performed with
  $N_{\rm max}=10$ and 50 discretized Berggren shells for the neutron
  $s_{1/2}$ partial wave.}
\label{fig:pot_lanc}
\end{center}
\end{figure}

\begin{figure}[htb]
\begin{center}
\includegraphics[scale=0.3]{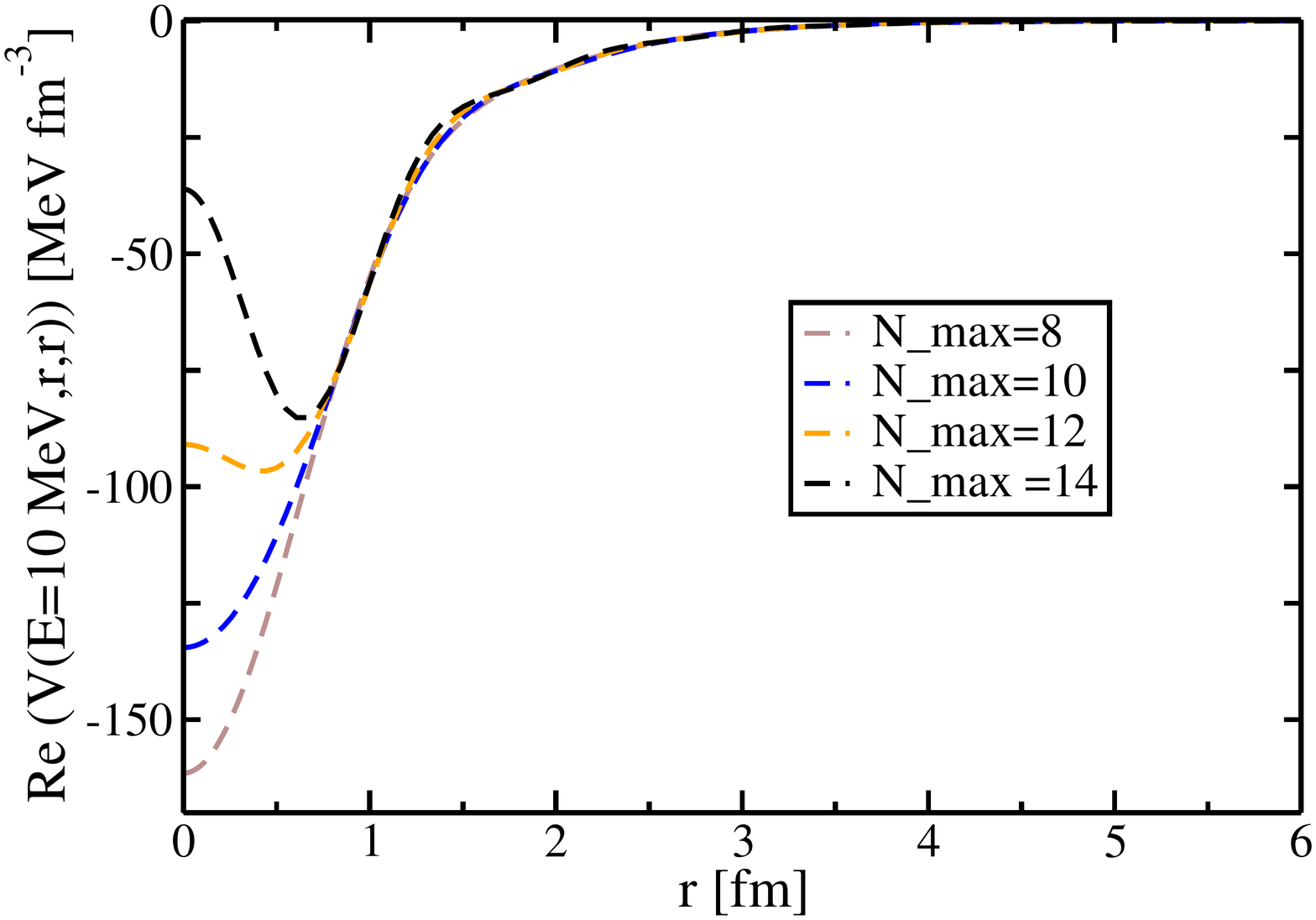}
\caption{Real part of the diagonal optical potential in the neutron
  $s$-wave at $E=10$ MeV. Results are shown for at ${ N_{\rm max}=8-14}$ and
  50 discretized shells in the $s_{1/2}$ partial wave}
\label{fig:pot_conv}
\end{center}
\end{figure}

We now illustrate the efficiency of the Lanczos method to calculate
the Green's function matrix elements (cf. Sec.~\ref{seclanc}) by
Figure~\ref{fig:pot_lanc} shows the convergence of the real part of
the radial (diagonal $r=r'$) $s$-wave optical potential as a function
of the number of Lanczos iterations $N_{\rm lanc}$.  Here, the
single-particle basis is based on a model space with
harmonic-oscillator shells up to $ N_{\rm max}=10$ and 50 discretized
Berggren $s_{1/2}$ shells.  We show results in Fig.~\ref{fig:pot_lanc}
for $E=10$~MeV.  After about 10 Lanczos iterations, the (diagonal)
potential quickly converges except in the vicinity of the origin
$r=0$ where the convergence is slower.  However, close to the origin
the $s$-wave scattering wavefunction $u(r)\sim r$, and the small
dependence on $N_{\rm lanc}$ will have a negligible impact on
observables.  As we will see later (cf  Fig.~\ref{fig:pot_conv}), the depth of the potential close to the origin 
depends on $N_{max}$ but again, due to the behavior of the scattering wave function in that region, this dependence will have 
a small impact on the results (see  Fig.~\ref{fig:pot_conv}). 

\begin{figure}[htb]
\begin{center}
\includegraphics[scale=0.3]{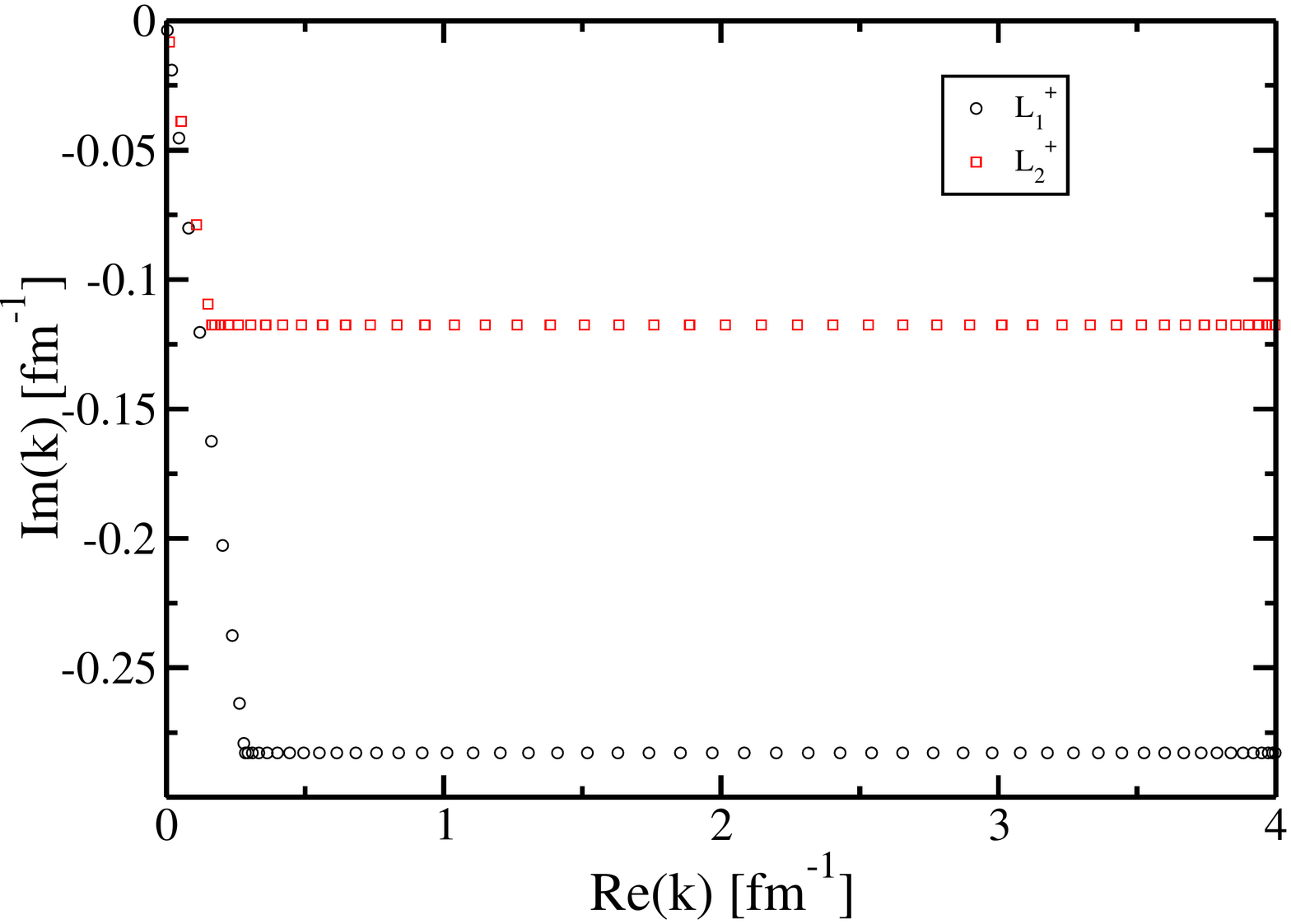}
\caption{k-plane contours $L^+_1,L^+_2$ used in the calculation of the $s$-wave optical potential in Fig.~\ref{fig:pot_real_dif_cont}.}
\label{fig:contk}
\end{center}
\end{figure}

\begin{figure}[htb]
\begin{center}
\includegraphics[scale=0.3]{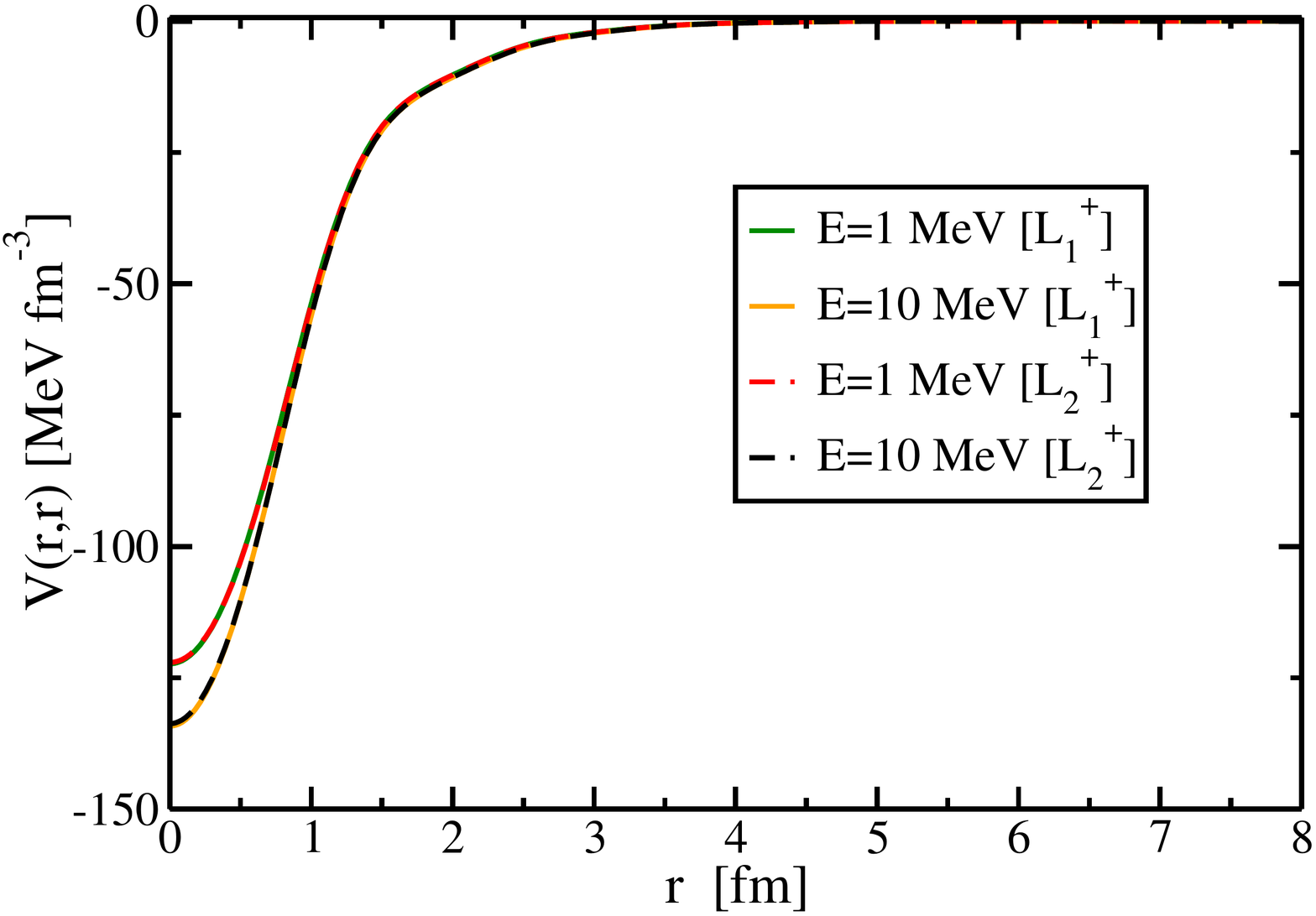}
\caption{Real part of the radial (diagonal) optical potential in the
  neutron $s$-wave at $E=1$ and $E=10$~MeV using two different
  contours $L^+_1,L^+_2$ for the single-particle neutron $s_{1/2}$
  shells,  in a model space with $N_{\rm max}=10$.}
\label{fig:pot_real_dif_cont}
\end{center}
\end{figure}

Results should be independent on the choice of the contour $L^+$ in
the complex momentum plane as long as its discretization is adequate
for the infrared scales under consideration~\cite{Furnstahl2012}.
Figure \ref{fig:pot_real_dif_cont} shows the real part of the radial
(diagonal) neutron $s$-wave potential at $E= 1$ and $E=10$~MeV using
two contours ${ L^+_1}$ and ${ L^+_2}$. 
 Both contours, shown in Fig.~\ref{fig:contk}, are defined by
two segments $[k_a,k_b] $ and $ [k_b,k_c]$ located on the fourth
quadrant of the complex momentum plane where $k_a$ is taken as the
origin. For the contour $L^+_1$, the segment $[k_a,k_b]$ has a norm of
${\rm 0.4 ~fm^{-1}}$, with an argument equal to $-\pi/4$ and $[k_b,k_c]$
is a horizontal segment with $\rm {Re(k_c)=4~fm^{-1}}$. For the
contour $L^+_2$, the segment $[k_a,k_b]$ has a norm of ${\rm 0.2
  ~fm^{-1}}$ and an angle equal to $-\pi/5$ and $[k_b,k_c]$ is a
horizontal segment with $\rm {Re(k_c)=4~fm^{-1}}$. We take 10 and 50
points on each segments for $L^+_1$, whereas we take 5 and 45 points
for the discretization of $L^+_2$,
respectively.
 Figure~\ref{fig:pot_real_dif_cont} shows that the
results are practically independent of the choice of the contour.
\\

\begin{figure}[htb]
\begin{center}
\includegraphics[scale=0.3]{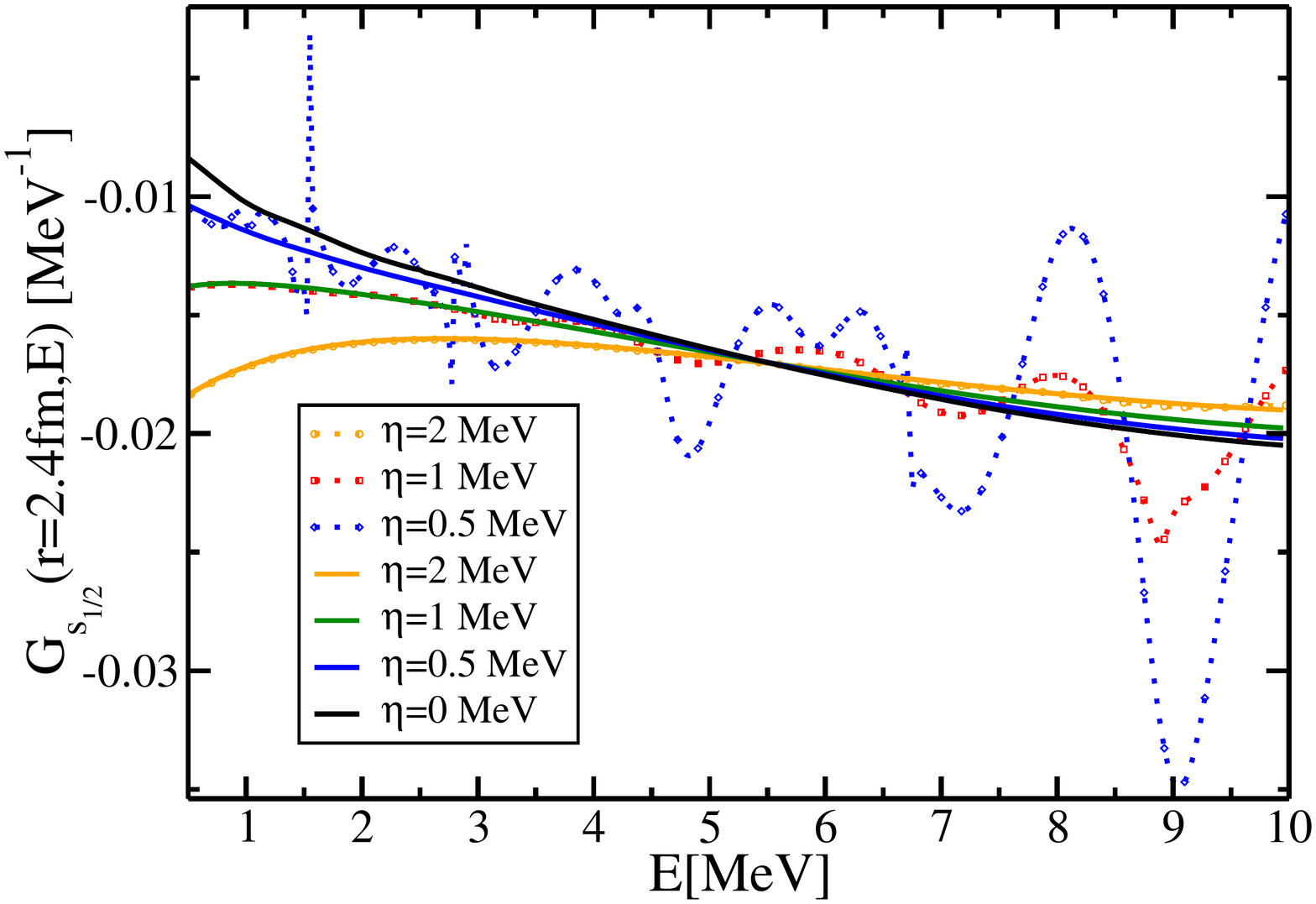}
\caption{Imaginary part of the neutron s-wave Green's function
  $G_{s1/2}(r,E)$ at $r=2.4$~fm (see text for details). The dashed
  lines correspond to calculations with a real HF basis whereas the
  full lines were obtained using a complex contour. Calculations were
  performed with ${N_{\rm max}=10}$ and 50 discretized shells in the
  ${s_{1/2}}$ partial wave.}
\label{fig:gf}
\end{center}
\end{figure}

In Fig.~\ref{fig:gf}, we illustrate the numerical stability of our
approach as $\eta \rightarrow 0$.  We show the imaginary part of the
(diagonal) $s$-wave Green's function $G_{s_{1/2}}(r,E)\equiv
G_{s_{1/2}}(r,r,E)$ using (i) a complex and (ii) real set of HF
orbitals for the $s_{1/2}$ neutron shells.  While the shown results
are for $r=2.4$~fm, we note that the qualitative behavior is
independent of the value of $r$. As expected (see Sec.~\ref{berg}),
for  $\eta$ values significantly larger than zero, both bases give the same results.
Let us first consider the real HF basis, corresponding to the dashed lines  in Fig.~ \ref{fig:gf}. For
$\eta=2$~MeV the results are smooth but, as $\eta$ decreases, the considerable oscillations appear, and for 
$\eta\sim0$ peaks with widths proportional to $\eta$ start to appear
near the Green's function poles, at real energies. If instead we use a complex single-particle basis (solid lines in Fig.~ \ref{fig:gf})
no such instability occurs  as $\eta \rightarrow 0$.

\begin{figure}[htb]
\begin{center}
\includegraphics[scale=0.3]{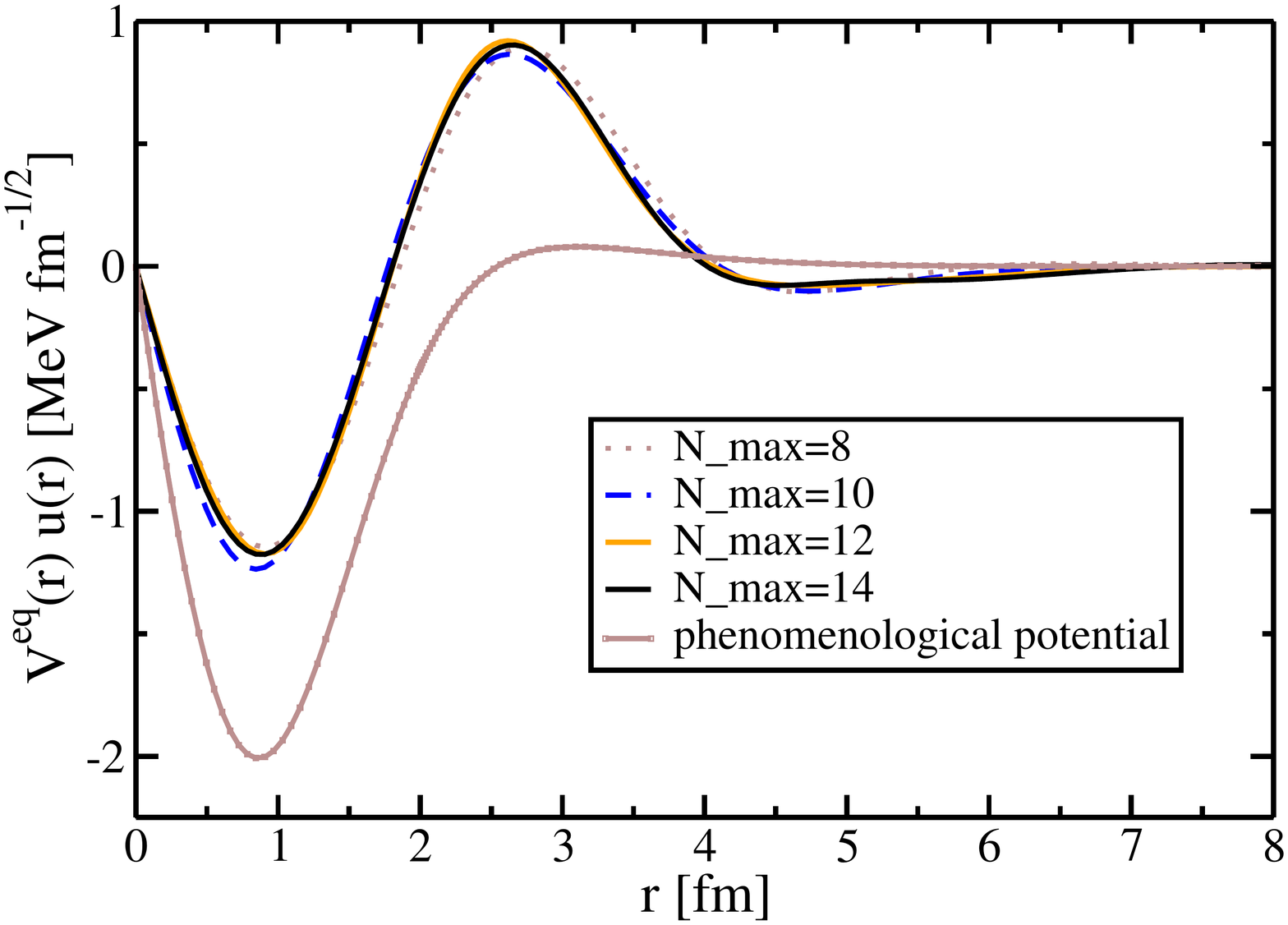}
\caption{Real part of $V_{\rm int}(r)$ in the neutron s-wave at E=10
  MeV. Results are shown for at ${ N_{\rm max}=8-14}$. For illustration
  purpose, we also show the results obtained with the phenomenological
  potential from Ref.~\cite{pheno}.}
\label{fig:pot_equi}
\end{center}
\end{figure}

Next, we show in Fig.~\ref{fig:pot_conv}, the convergence of the real
part of the (diagonal) the $s$-wave optical potential as the size of
the model space increases from ${N_{\rm max}=8}$ to $14$. Results are
shown for $E=10$~MeV and, in all cases, 50 discretized shells are used for the Berggren basis in the $s$-wave,  and $\eta=0$.  For $N_{max}=10$ and $E=10$~MeV, the
results agree with those shown in Fig.~\ref{fig:pot_real_dif_cont}.
Convergence is achieved for $N_{\rm max}=14$ for $ r \geq 1$~fm. For
small values of $r$, the optical potential depends on $N_{\rm max}$.
This is understandable because short-range physics gets better resolved as
the model space increases., and thus convergence becomes harder. Again we note that in this region the
scattering wave function $u(r) \propto r$ and the dependence of the
potential on ${N_{\rm max}}$ does not impact observables. To
demonstrate this point, Fig.~\ref{fig:pot_equi} shows the integrated
quantity
\begin{eqnarray}
V_{\rm int}(r) \equiv r\int dr' r' V(r,r')u(r')=V^{\rm eq}(r)u(r) .
\end{eqnarray}

The potential $V_{\rm int}(r)$ can be viewed as the local equivalent
potential $V^{\rm eq}(r)$ multiplied by the scattering wave
function, and corresponds to the source term in the one-body optical-model-type Schr\"odinger equation. 
The variations of the optical potential with the model space for small
values of $r$ do not impact the behavior of $V_{\rm int}(r)$. For
illustration, Fig.~\ref{fig:pot_equi} also shows a result for $V_{\rm
  int}(r)$ obtained using a phenomenological potential based on a
Woods-Saxon form factor \cite{pheno}.

\begin{figure}[htb]
\begin{center}
\hspace*{-3cm}\includegraphics[scale=0.5]{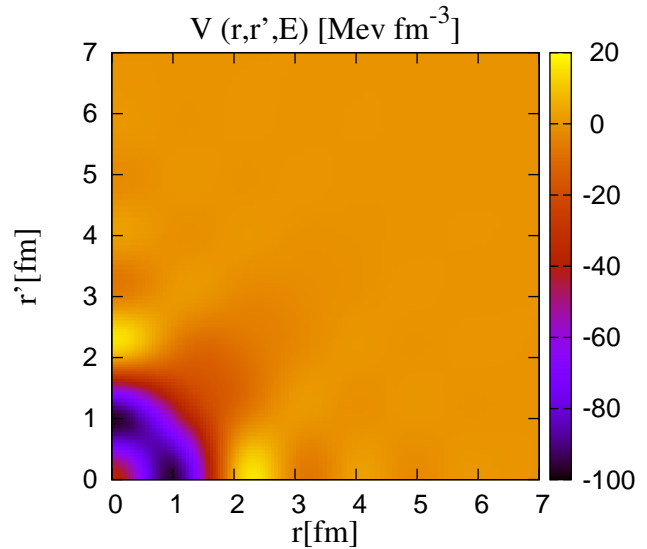}
\caption{Contour plot of the real part of the neutron s-wave potential
  $V(r,r',E)$ for ${ N_{\rm max}=14}$ and 50 discretized $s$-wave shells at
  $E=10$~MeV.}
\label{fig:s_contour}
\end{center}
\end{figure}

\begin{figure}[htb]
\begin{center}
\includegraphics[scale=0.3]{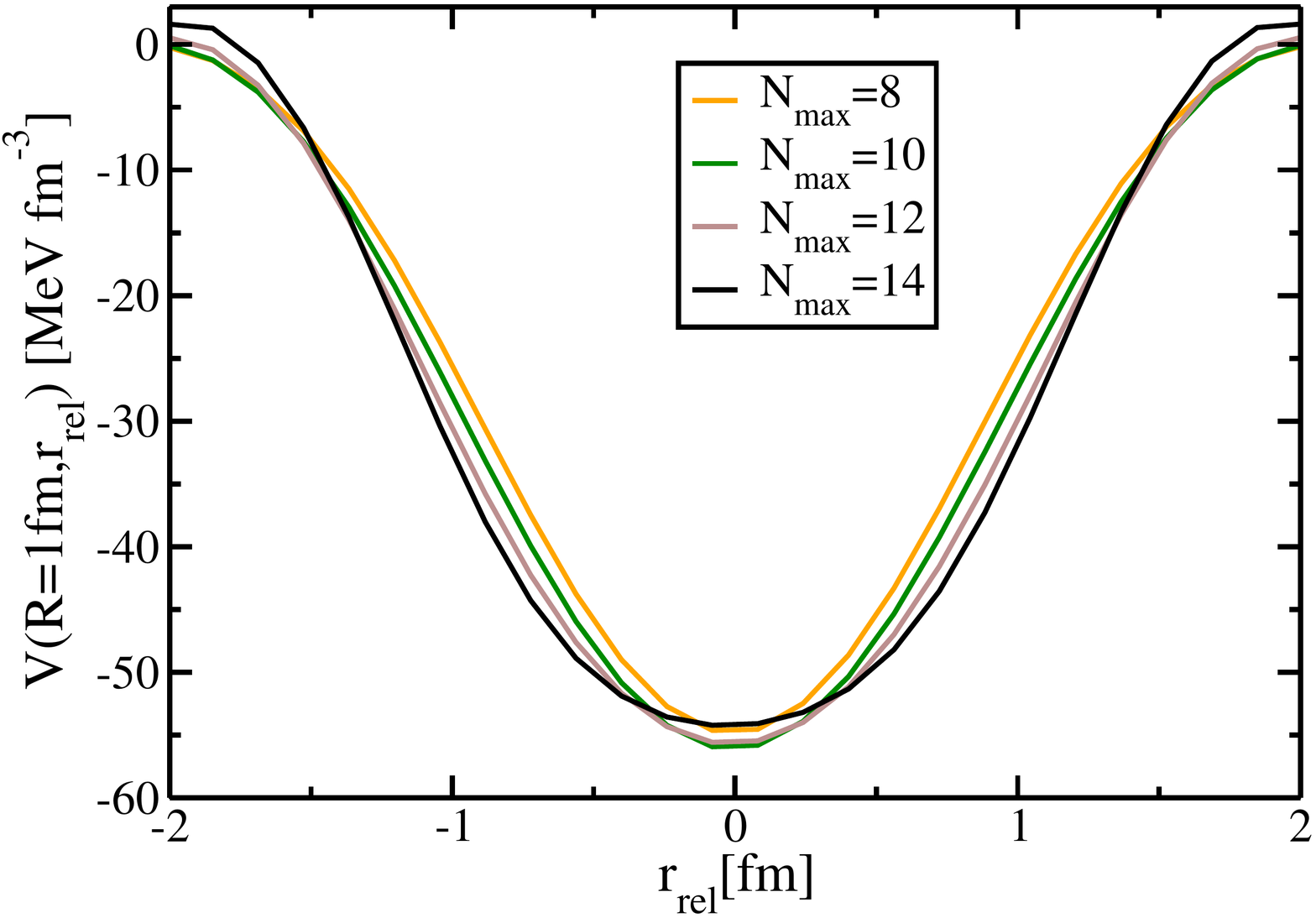}
\caption{Neutron $s$-wave optical potential at E=10 MeV plotted as
  $V(R+r_{\rm rel}/2,R-r_{\rm rel}/2)$ at fixed $R=1/2$~fm.  Here
  $N_{max}=14$ and 50 discretized $s$-wave shells are included in the
  single-particle basis.}
\label{fig:counter_diag_s_10_MeV}
\end{center}
\end{figure}

So far, we  have only presented results for the diagonal part of the
optical potential. Figure~\ref{fig:s_contour} shows a contour plot for
the nonlocal neutron $s$-wave optical potential. Introducing the
relative coordinate $r_{\rm rel}=r-r'$ and the center-of-mass
coordinate $R=(r+r')/2$ we plot the optical potential as a function of
$r_{\rm rel}$ at fixed $R=1$~fm in
Fig.~\ref{fig:counter_diag_s_10_MeV}. We can see that the full width
at half maximum is about 2.2~fm. Clearly, this potential is very
different from a model of a Dirac $delta$ function in $r_{\rm rel}$ and exemplifies the degree of
nonlocality which is predicted microscopically.  We note that due to the non-Hermitian nature of the
coupled-cluster method, the potential $V(r,r')$ is slightly
non-symmetric in $r$ and $r'$, and as a consequence $V(R,r_{\rm rel})$
is not quite an even function of $r_{\rm rel}$.  In
Figs.~\ref{fig:s_contour}, and \ref{fig:counter_diag_s_10_MeV} the
energy is $E=10$ MeV and results were obtained for ${ N_{\rm max}=14}$ and 50 discretized shells for the $s$-wave  along a contour in the complex plane.

\begin{figure}[htb]
\begin{center} 
\hspace*{-3cm}\includegraphics[scale=0.5]{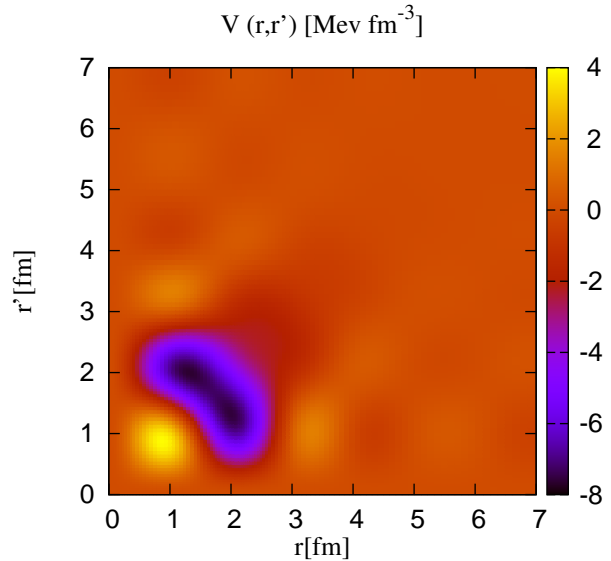}
\caption{Contour plot of the real part of the neutron $ d_{3/2}$-wave
  potential for ${N_{\rm max}=14}$ at $E=10$~MeV.}
\label{fig:d_contour}
\end{center}
\end{figure}

Calculations of the optical potential in other partial waves follow
along the same lines.  For illustration, we show a contour plot of the ${
  d_{3/2}}$-wave potential in Fig.~\ref{fig:d_contour}. Results are
shown for $E=10$~MeV at ${ N_{\rm max}=14}$ and 50 discretized shells for the $d_{3/2}$-wave
 along a complex contour. As in other cases, we take the limiting value $\eta =0$.
\\

\begin{figure}[htb]
\begin{center}
\includegraphics[scale=0.3]{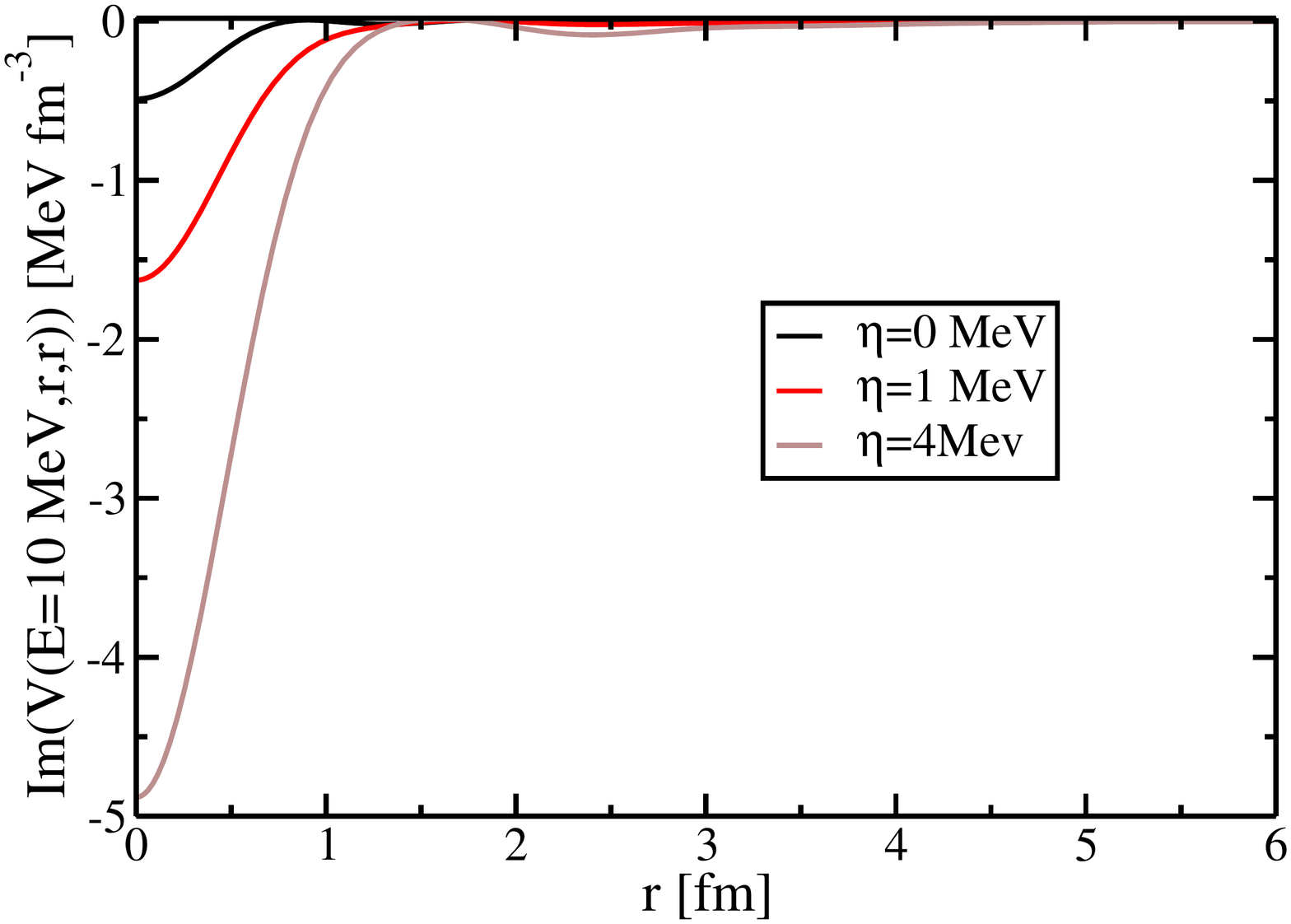}
\caption{Imaginary part of the radial (diagonal) optical potential in
  the neutron $s$-wave at $E=10$~MeV.  Calculations were performed at
  ${ N_{\rm max}=10}$ with 50 $s_{1/2}$ discretized shells. Results
  are shown for several values of $\eta$.}
\label{fig:pot_imag}
\end{center}
\end{figure}
We finally turn to the imaginary part of the optical potential.  The
imaginary part describes the loss of flux due to inelastic processes.
For most nuclei, and particularly for heavier systems, there are many compound-nucleus resonances above
the particle threshold, and absorption is known to be significant. 
Our results for the imaginary part of the potential, along the diagonal $r=r'$ are shown in  Fig.~\ref{fig:pot_imag} for the neutron $ s_{1/2}$ wave at $E=10$~MeV. The model space consists of $N_{max}=10$ and 50 discretized shells for the s-wave.
We consider various values of $\eta$. In the limit $\eta=0$, the imaginary part of the potential is very small, and this is true for the whole range of energies up to $E=10$~MeV.
As one can see in Fig.~\ref{fig:pot_imag}, as $\eta$ decreases to zero,
the imaginary part also decreases and becomes very small for $\eta=0$.
We observed the same qualitative behavior for all other considered
partial wave, up to  $d_{5/2}$, a result that does not change when the model space increases.

\begin{figure}[htb]
\begin{center}
\includegraphics[scale=0.3]{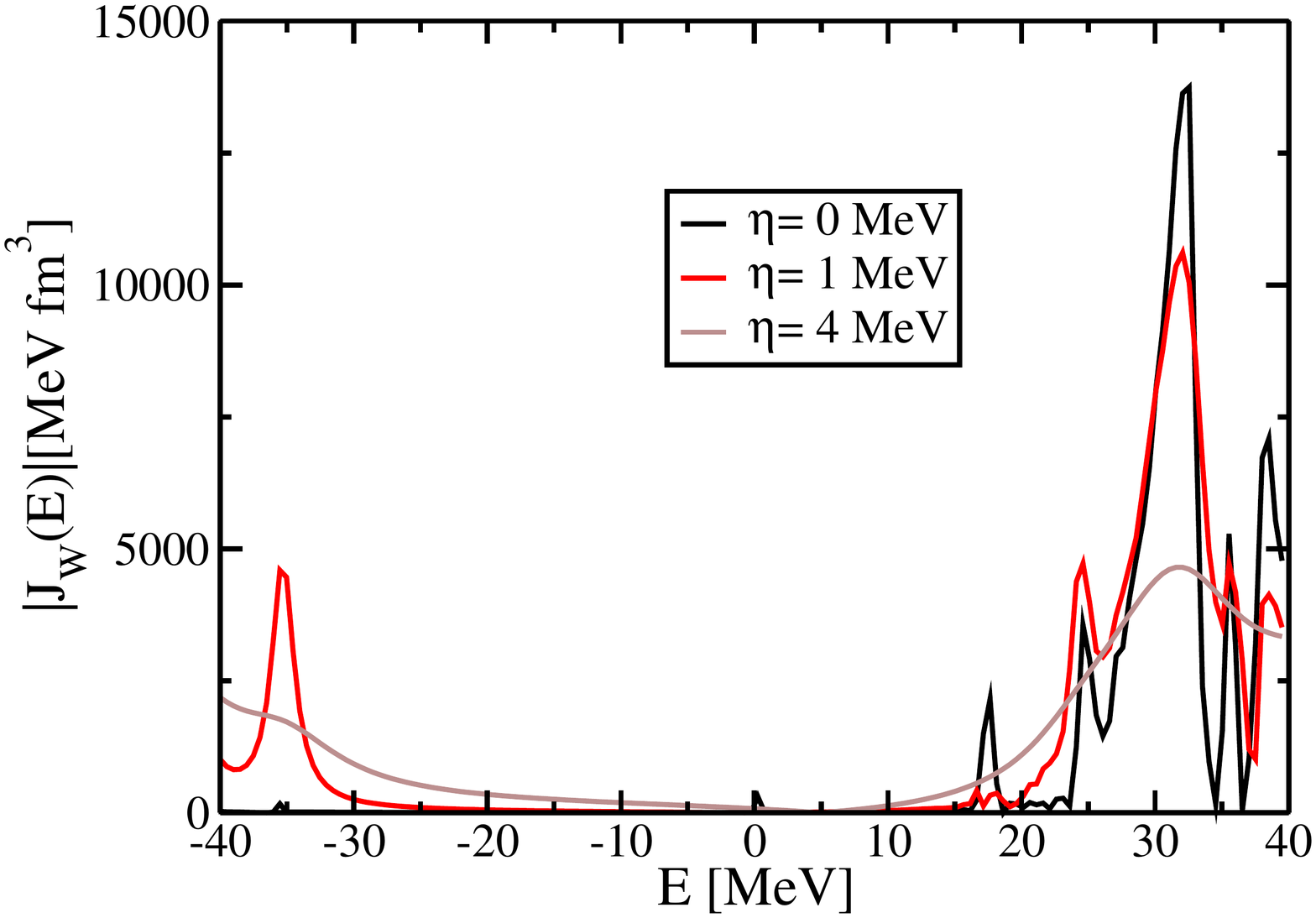}
\caption{Neutron $s$-wave imaginary volume integral $J_{W}(E)$ for
  several values of $\eta$. Calculations were performed at
  $N_{\rm max}=14$ with 50  discretized $s_{1/2}$ shells.}
\label{fig:Jw}
\end{center}
\end{figure}

To further illustrate our difficulties with the imaginary part, we plot in 
Fig.~\ref{fig:Jw} the imaginary volume integral $J^l_{W}$
\begin{eqnarray}
J^l_{W}=4\pi\int dr r^2 \int dr r'^2 {\rm Im} \Sigma'_l(r,r';E)
\end{eqnarray} 
for the optical potential in the $s$-wave, taking a model space with $N_{\rm max}=14$ and 50 discretized  s-waves. 

In order to understand these results, we recall that the compound states that contribute to the flux removal from the elastic channel consist of a high number of particle-hole
excitations and are usually described by stochastic approaches~\cite{mitchell2010}.  However, the coupled-cluster approach
to the optical potential presented in this paper employs only
$1p$-$1h$ and $2p$-$2h$ excitations and is thus limited to absorption
on resonant states that are dominated by $1p$-$1h$ excitations. In our
example of scattering off $^{16}$O, its $J^\pi=3^-$ state (at about
6~MeV of excitation energy) is thought to be of $1p$-$1h$ structure.
With the NNLO$_{\rm opt}$ interaction, we computed this state using
EOM-CCSD and found it at about 10~MeV of excitation. Another relevant excited state in $^{16}$O is the first excited $0^+$ state also at $\approx6$ MeV, which is known to have a strong $4p-4h$ configuration. In our coupled cluster calculations this state is above $10$ MeV. In fact, there are no other excited states below 10~MeV. 
In general, positive parity states of $^{16}$O are dominated by $2p$-$2h$
excitations, and are therefore not well described in EOM-CCSD. Thus, from this analysis, we conclude that it is not possible to
 produce significant absorption at low-energies for neutron
scattering on $^{16}$O due to the employed low-order cluster
truncations in our EOM-CCSD and PA/PR-EOM CCSD approximations. 

One path forward is to introduce a phenomenological and energy dependent width in the Green's
function, to account for higher-order correlations such as $3p$-$2h$
and $2p$-$3h$ not included in PA/PR-EOM CCSD \cite{dickhoff2016}. As shown in Fig.\ref{fig:pot_imag},
this will increase the absorption at lower energies.
This would also allow to account for collective states 
which may exist in nature and which cannot be reproduced in the coupled cluster approach at the CCSD level. 

\begin{figure}[htb]
\begin{center}
\includegraphics[scale=0.3]{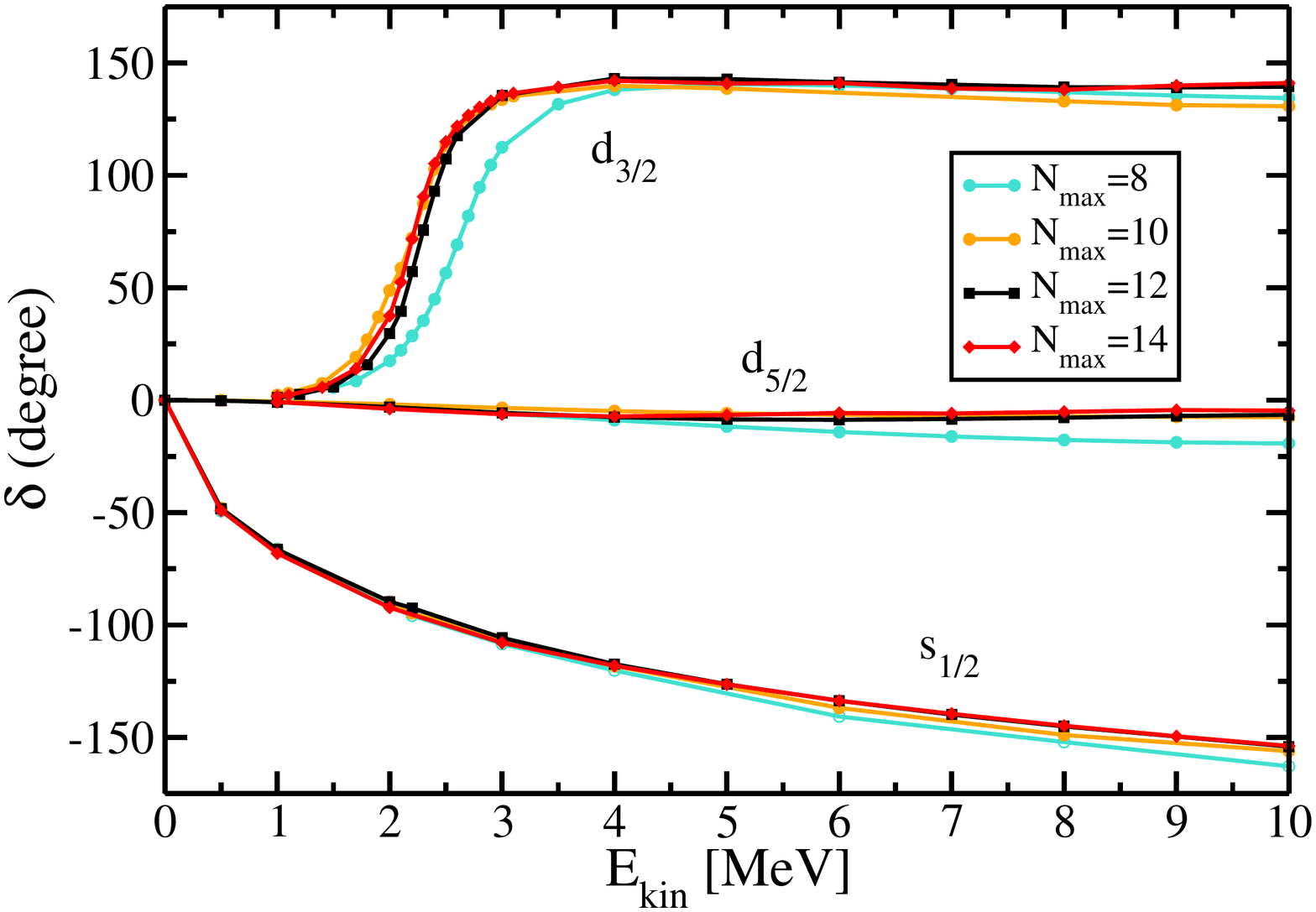}
\caption{Elastic-scattering phase shifts in the neutron $s$ and $d$
  waves as a function of $N_{\rm max}$. In all cases 50 discretized
  Berggren shells are included.}
\label{fig:phase_shift}
\end{center}
\end{figure}

Finally we show, in Fig.~\ref{fig:phase_shift}, the neutron elastic
scattering phase shift obtained with the optical potential in the $s$
and $d$ partial waves, as a function of the model space
\footnote{In principle, the phase shift should be
obtained by solving the Schr\"odinger Eq.~\ref{schro} in the relative
coordinate, with the reduced mass $\mu_{{n-^{16}{\rm O}}}$ of the
$n-^{16}$O system. However, with the optical potential being calculated in
the laboratory frame (the Hamiltonian $H$ (\ref{hami}) is
defined in the laboratory) a correction to the reduced mass is needed.
This correction is such that the reduced mass $\mu'$ used to solve the
Schr\"odinger Eq.~(\ref{schro}) is $1/\mu'=(1-1/A)/m$ (cf
Eq.~\ref{hami}). Doing so, the bound states of the optical potential
in the $d_{5/2}$ and $s_{1/2}$ partial waves correspond to
respectively, the gs and first excited state in $^{17}$O obtained with
the PA-EOM CCSD method.}. 
We want to emphasize here that calculations for higher partial waves proceed similarly and  are straightforward. 
 We find that, for $N_{\rm max}=14$  all calculated phase shifts have
converged (all calculations here are done with 50 discretized shells). 
 The sharp rise of the phase shift in the $d_{3/2}$ partial
wave is the standard signature of the resonance $J^{\pi}=3/2^{+}$ in
$^{17}$O, which is numerically predicted to be at  $E=2.26-i0.12$~MeV from our PA-EOM CCSD calculations (see Table~\ref{tabres}).

\section{Conclusions} \label{conclusion} 
 We constructed microscopic nuclear optical potentials by combining the
Green's function approach with the coupled-cluster method.  For the
computation of the Green's function, we used an analytical
continuation in the complex energy plane, based on a Berggren
basis. Using the Lanczos method, we expressed the Green's function as
a continued fraction. The computational cost of a single Lanczos
iteration is similar to that of a PA-EOM-CCSD calculation,
i.e. polynomial in system size, and thus affordable. The convergence
with the number of Lanczos iterations was demonstrated.  The Dyson
equation was then inverted to obtain the optical potential.

In the coupled-cluster
singles and doubles approximation, the optical potential and the
neutron elastic scattering phase shifts on $^{16}{\rm O}$ converge well
with respect to the size of the single-particle basis, for the low partial waves.  
The predicted optical potential has a strong nonlocality that is not Gaussian. In addition, we found an
almost vanishing imaginary part of the potential for scattering
energies below  10~MeV. This lack of an absorptive component was
attributed to neglected higher-order correlations in the employed
coupled-cluster methods.  

In the future, we plan to  update the NN force currently used, to one that is able to reproduce charge radii of heavier systems. We also plan to include three-nucleon forces in the coupled-cluster calculations of the Green's functions, as well as higher-order correlations in the employed coupled-cluster methods. We expect this will produce an increase in the imaginary part of the derived optical potential.
Once these improvements are in place, this work can be extended to other systems (the limitations being the computational cost associated with the CC calculations) and to other reaction channels such as transfer, capture, breakup and charge-exchange. Systematic studies involving heavier nuclei and consistent calculations along isotopic chains will provide critical information on how to extrapolate the optical potential to unknown regions of the nuclear chart.

\begin{acknowledgments}
We acknowledge beneficial discussions with Carlo Barbieri, Willem Dickhoff, Charlotte
Elster, Gr\'egory Potel and R. C. Johnson.  This work was supported by the Office of
Nuclear Physics, U.S. Department of Energy under contracts 
DE-FG02-96ER40963, DE-FG52- 08NA28552 (RIBSS Center) and DE-SC0008499 (NUCLEI SciDAC collaboration), and the
Field Work Proposal ERKBP57 at Oak Ridge National Laboratory (ORNL).
We also acknowledge the support of the National Science Foundation under Grants
No. PHY-1520929 and PHY-1403906.
Computer time was provided by the Institute for Cyber-
Enabled Research at Michigan State University 
and the Innovative and Novel Computational Impact on Theory and Experiment (INCITE) program. This research used
resources of the Oak Ridge Leadership Computing Facility located at
ORNL, which is supported by the Office of Science of the Department of
Energy under Contract No.  DE-AC05-00OR22725.
\end{acknowledgments}


%

\end{document}